\documentclass[lettersize,journal]{IEEEtran}

\usepackage{amsmath,amsfonts}
\usepackage{algorithmic}
\usepackage{algorithm}
\usepackage{array}
\usepackage[caption=false,font=normalsize,labelfont=scriptsize,textfont=scriptsize]{subfig}
\usepackage{url}
\usepackage{bm}
\usepackage{verbatim}
\usepackage{graphicx}
\usepackage{cite}
\usepackage{amssymb}
\usepackage{threeparttable}
\begin{document}
\title{Stacked Intelligent Metasurfaces for \\ Task-Oriented Semantic Communications}
\author{Guojun Huang, Jiancheng An, \IEEEmembership{Member, IEEE}, Zhaohui Yang, \IEEEmembership{Member, IEEE}, Lu Gan, \IEEEmembership{Member, IEEE},\\Mehdi Bennis, \IEEEmembership{Fellow, IEEE}, and Mérouane Debbah, \IEEEmembership{Fellow, IEEE}
\thanks{This work is partially supported by the Sichuan Science and Technology Program under Grant 2023YFSY0008 and 2023YFG0291, and is partially supported by the National Natural Science Foundation of China (Grant number 62471096). G. Huang and L. Gan are with the School of Information and Communication Engineering, University of Electronic Science and Technology of China (UESTC), Chengdu, Sichuan, 611731, China. L. Gan is also with the Yibin Institute of UESTC, Yibin 644000, China (e-mail: guojun$\_$huang2000@sina.com; ganlu@uestc.edu.cn). J. An is with the School of Electrical and Electronics Engineering, Nanyang Technological University, Singapore 639798 (e-mail: jiancheng\_an@163.com). Z. Yang is with the College of Information Science and Electronic Engineering, Zhejiang University, Hangzhou, Zhejiang 310027, China. (e-mail: yang\_zhaohui@zju.edu.cn). M. Bennis is with the Centre for Wireless Communications, University of Oulu, 90570 Oulu, Finland (e-mail: mehdi.bennis@oulu.fi). M. Debbah is with the Center for 6G Technology, Khalifa University of Science and Technology, P O Box 127788, Abu Dhabi, United Arab Emirates (e-mail: merouane.debbah@ku.ac.ae).}}
\markboth{Draft}{Draft}
\maketitle
\begin{abstract}
Semantic communication (SemCom) leveraging advanced deep learning (DL) technologies enhances the efficiency and reliability of information transmission. Emerging stacked intelligent metasurface (SIM) with an electromagnetic neural network (EMNN) architecture enables complex computations at the speed of light. In this letter, we introduce an innovative SIM-aided SemCom system for image recognition tasks, where a SIM is positioned in front of the transmitting antenna. In contrast to conventional communication systems that transmit modulated signals carrying the image information or compressed semantic information, the carrier EM wave is directly transmitted from the source. The input layer of the SIM performs source encoding, while the remaining multi-layer architecture constitutes an EMNN for semantic encoding, transforming signals into a unique beam towards a receiving antenna corresponding to the image class. Remarkably, both the source and semantic encoding occur naturally as the EM waves propagate through the SIM. At the receiver, the image is recognized by probing the received signal magnitude across the receiving array. To this end, we utilize an efficient mini-batch gradient descent algorithm to train the transmission coefficients of SIM's meta-atoms to learn the semantic representation of the image. Extensive numerical results verify the effectiveness of utilizing the SIM-based EMNN for image recognition task-oriented SemComs, achieving more than 90\% recognition accuracy.
\end{abstract}


\section{Introduction}
\IEEEPARstart{C}{onventional} content-centric communication paradigms fail to satisfy the stringent quality-of-service requirements of next-generation wireless networks, particularly in terms of ultra-reliability, low latency, and high data rates \cite{wang2023road}. As a result, there is a growing interest in intent-driven networks, which triggers the recent research surge in semantic communications (SemComs) \cite{yang2022semantic, Yang2024Secure}. Relying on advanced artificial intelligence technologies and powerful computing capability, SemCom enables learning the semantic representations of various types of sources. For instance, the authors of \cite{ITSP_xie2021deep} proposed a novel text SemCom system, \textit{DeepSC}, to recover the meaning of sentences at the receiver with the minimum semantic errors. Based on this, several variants of \textit{DeepSC} and advanced deep learning (DL)-driven SemCom systems have been developed \cite{Selected_Areas_xie2023semantic, qin2024computing}. Nevertheless, the joint optimization of the semantic and channel encoders in these SemCom systems is typically time-consuming, which undoubtedly increases the training costs and deployment challenges for edges \cite{qin2024computing}. Additionally, the inference process would suffer from high computational complexity as the number of nodes increases.

Recently, stacked intelligent metasurface (SIM) emerged as a promising technique for realizing fast and efficient wave-domain computation. Specifically, a SIM is a closed vacuum container with multiple closely stacked metasurface layers, where each metasurface consists of numerous low-cost meta-atoms whose electromagnetic (EM) response can be flexibly tuned by an intelligent controller, e.g., a field programmable gate array (FPGA) \cite{An2024Stacked, TCOM_2024_Yu_Environment}. As a result, SIM forms a programmable electromagnetic neural network (EMNN) in the EM domain, with signal processing implemented at the speed of light \cite{n_yao2020protonic, an2023Stacked, yao2024channel, Papaza2024Achievable}. Recently, researchers have utilized SIMs to perform various computing tasks in wireless communication and sensing scenarios, such as multiuser precoding \cite{liu2024drl, TWC_2024_Anastasios_Achievable, WC_2024_An_Near, arXiv_2024_Shi_Joint, APWCS_2024_Li_Stacked} and direction-of-arrival (DOA) estimation \cite{an2024two, WCL_2024_Niu_Stacked}. Nevertheless, the potential of utilizing SIM for SemComs has not hitherto been investigated.

Against this background, we examine an image recognition task-oriented SemCom system leveraging a SIM. Compared to conventional SemCom systems sequentially encoding and transmitting semantic information, the wave-based computing paradigm enables automatic and efficient source and semantic encoding as the EM waves propagate through the SIM, greatly reducing energy consumption and processing latency. Specifically, in the proposed system, a SIM is positioned in front of the transmitting antenna, which only transmits the carrier EM waves. The input layer of the SIM performs image source encoding, while the remaining multi-layer architecture constitutes an EMNN for image semantic encoding, transforming the signals passing through the input layer of the SIM into a unique beam towards an antenna corresponding to the image class. As a result, the receiver recognizes the image by probing the signal magnitude across the receiving array. To this end, we utilize an efficient mini-batch gradient descent algorithm to train the transmission coefficients of meta-atoms in the SIM. Finally, extensive simulation results verify the effectiveness of utilizing a SIM to realize the task-oriented SemComs.

\begin{figure}[!t]
\centering
\includegraphics[width=8.8cm]{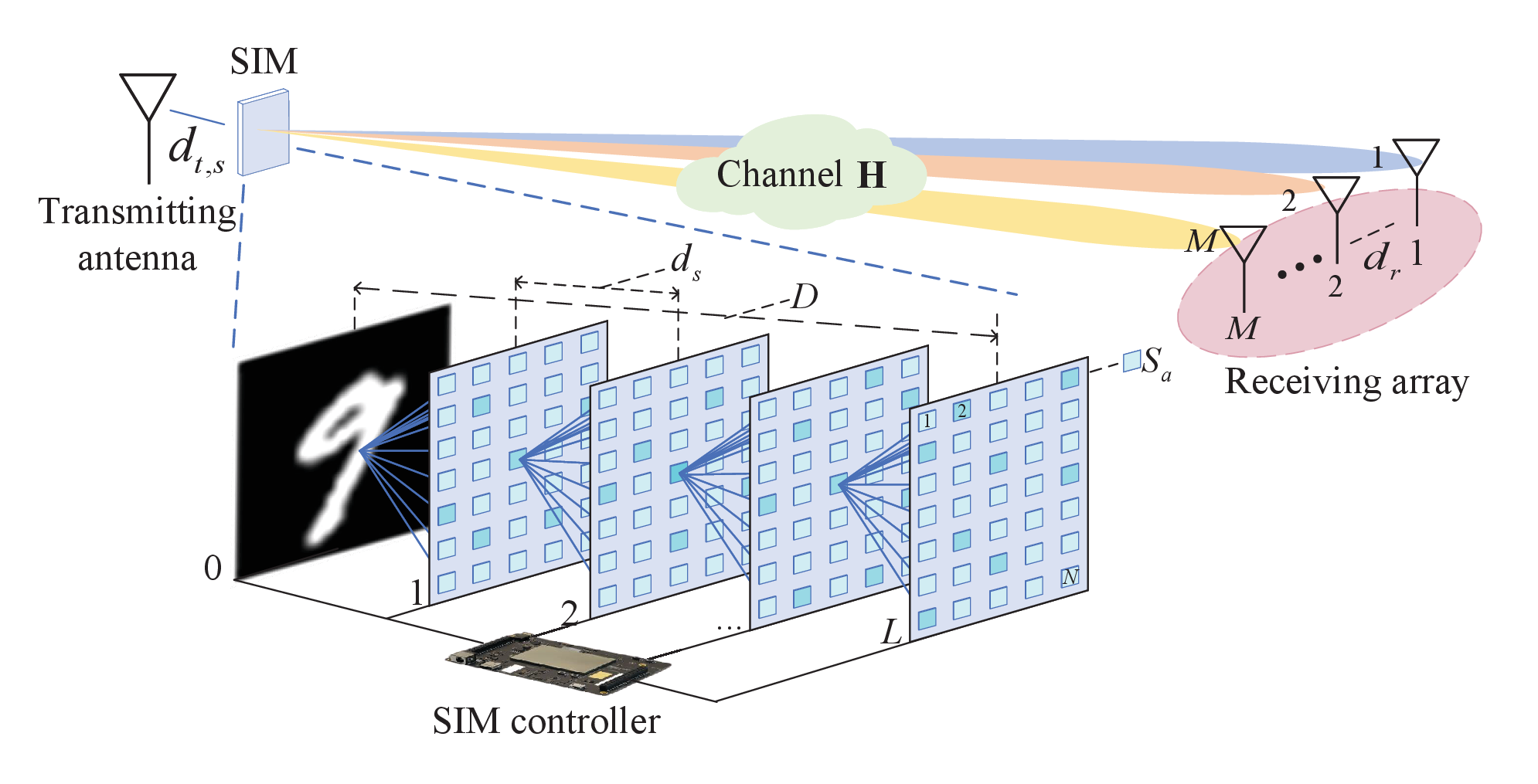}
\caption{A SIM-aided image recognition task-oriented SemCom system, where we consider $M$ image classes.}
\label{fig_1}
\end{figure} 
\section{System Model}
In this section, we introduce the SIM-aided SemCom system for image recognition. As shown in Fig. 1, we consider a system consisting of a single-antenna transmitter and a receiver equipped with $M$ antennas, which correspond to the number of image classes. Let $\mathcal{M} = \left\{1, 2, \ldots, M\right\}$ denote the set of indexes of antennas. More specifically, the transmitting antenna emits the unmodulated radio wave, and the SIM performs source and semantic coding of the image to direct the transmissive EM waves towards the corresponding antenna at the receiver.
\subsection{Transmitter}
Specifically, the transmitter consists of two components: i) a single transmitting antenna, and ii) a SIM in front of it.
\subsubsection{Transmitting Antenna}
In contrast to conventional communication systems, the transmitting antenna only radiates the carrier EM waves. Let $\sqrt{p_{t}}s$ denote the transmit signal, where $p_{t}$ represents the transmit power, and $s$ is the normalized EM signal\footnote{Note that the receiver recognizes the image class by probing the amplitude of the EM signal, which means that the phase of $s$ has no impact on the recognition performance.}, satisfying $|s|^{2} = 1$.

\subsubsection{SIM}
Furthermore, a SIM is positioned in front of the transmitting antenna to perform complex computations for source and semantic coding. Let $(L+1)$ denote the number of layers, while the corresponding set is represented by $\mathcal{L} = \left \{0, 1, 2, \ldots, L \right \}$. Let $D$ and $d_s$ denote the thickness of the SIM, and the spacing between adjacent layers, respectively. Thus, we have $d_{s} = D / L$. Moreover, let $N$ and $\mathcal{N} = \left \{1, 2, \ldots, N \right \}$ denote the number of meta-atoms on each layer as well as the corresponding set. Each metasurface layer is modeled as a uniform planar array, where meta-atoms are identically arranged in a square configuration. Specifically, the numbers of meta-atoms in rows and columns are denoted by $N_{\mathrm{row}}$ and $N_{\mathrm{col}}$, respectively. Thus, we have $N = N_{\mathrm{row}} N_{\mathrm{col}}$. Let $z^{l}_{n} = a^{l}_{n} e^{j \phi^{l}_{n}}$ denote the transmission coefficient of the $n$-th meta-atom on the $l$-th metasurface layer \cite{liu2022programmable}, where $a^{l}_{n} \in [0,1)$ and $\phi^{l}_{n} \in [0, 2\pi)$ represent the corresponding amplitude and phase shift, respectively. The transmission coefficient matrix of the $l$-th metasurface layer is given by $\mathbf{Z}^{l}= \mathbf{A}^{l} \bm{\Phi}^{l}=\mathrm{diag}\left(z^{l}_{1},z^{l}_{2},\ldots,z^{l}_{N}\right)\in\mathbb{C}^{N\times N}$, where $\mathbf{A}^{l} = \mathrm{diag}\left(\mathbf{a}^{l}\right) = \mathrm{diag} \left(a^{l}_{1}, a^{l}_{2}, \ldots, a^{l}_{N}\right) \in \mathbb{R}^{N \times N}$, $\bm{\Phi}^{l} = \mathrm{diag}\left(e^{\bm{\phi}^{l}}\right) = \mathrm{diag} \left(e^{j\phi^{l}_{1}}, e^{j\phi^{l}_{2}}, \ldots, e^{j \phi^{l}_{N}}\right) \in \mathbb{C}^{N \times N}$. 

The EM propagation coefficients between adjacent layers in SIM can be approximately characterized using the Rayleigh-Sommerfeld diffraction equation \cite{lin2018all}. Let $\mathbf{W}^{l} \in \mathbb{C}^{N \times N}$ denote the propagation matrix from the $\left(l-1\right)$-th layer to the $l$-th layer, with $w^{l}_{n, n^{*}}$ representing the entry on the $n$-th row and $n^{*}$-th column. Specifically, $w^{l}_{n, n^{*}}$ denotes the EM propagation coefficient from the $n^{*}$-th meta-atom on the $\left(l-1\right)$-th layer to the $n$-th meta-atom on the $l$-th metasurface layer, which is defined by
\begin{align}
 w^{l}_{n, n^{*}} & = f_{w}\left(d_{s}, d_{n,n^{*}}^{l}, S_{a}, \lambda\right) \notag\\ & = \frac{d_{s}S_{a}}{\left({d_{n,n^{*}}^{l}}\right)^{2}}\left(\frac{1}{2\pi d_{n,n^{*}}^{l}}-j\frac{1}{\lambda}\right)e^{j2\pi d_{n,n^{*}}^{l}},
\end{align}
for $l\in\mathcal{L}/\{0\}$, $n^{*}, n\in\mathcal{N}$, where $S_{a}$ is the area of each meta-atom, $d^{l}_{n,n^{*}}$ denotes the corresponding propagation distance between the two meta-atoms, $\lambda$ denotes the radio wavelength. Moreover, since the SIM is closely placed in front of the transmitter, the propagation channel from the transmitting antenna to the input layer of the SIM is characterized by $\mathbf{w}^{0} = f_{w}\left(d_{t,s}, d^{0}_{n}, S_{a}, \lambda\right) \in \mathbb{C}^{N}$ \cite{an2023Stacked}, where $d_{t,s}$ and $d^{0}_{n}$ denote the horizontal distance from the transmitting antenna to the input layer of the SIM and the distance from the transmitter to the $n$-th meta-atom on the input layer, respectively.

\paragraph{Source Encoding Layer} As illustrated in Fig. 1, the input layer of the SIM serves as a digital-to-analog converter to transform the pixel matrix of the source image into the transmission coefficients of the meta-atoms on the input layer\footnote{For the sake of brevity, we consider only greyscale images in this letter. For color images, each pixel matrix of three channels can be preprocessed for fully leveraging the amplitude and phase components of the transmission coefficients of meta-atoms on the input layer.}. Specifically, the image is first preprocessed to match the size of the input layer $\left(N_{\mathrm{row}} \times N_{\mathrm{col}}\right)$ by employing appropriate extraction or interpolation techniques \cite{jahne2005digital}, and then the transmission coefficients of the source encoding layer $\left(z^{0}_{1},z^{0}_{2},\ldots,z^{0}_{N}\right)$ are configured to the corresponding value of the pixel matrix. The amplitude and phase of the EM waves passing through the source encoding layer are determined by the product of the incident signal on the source coding layer, i.e., $\sqrt{p_{t}}s \mathbf{w}^{0}$ and the EM response imposed by the input layer, i.e., $\mathbf{Z}^{0}$. Therefore, the source encoded signal $\mathbf{x} \in \mathbb{C}^{N}$ radiated from the input metasurface layer can be expressed as
\begin{align}
 \mathbf{x} = \sqrt{p_{t}}s \mathbf{Z}^{0} \mathbf{w}^{0}.
\end{align}
\paragraph{Semantic Encoding Layers} The rest layers of the SIM construct an EMNN, which serves as a semantic encoder, aiming to transform $\mathbf{x}$ to a unique beam towards the receiving antenna corresponding to its image class. According to the Huygens–Fresnel principle \cite{liu2022programmable}, the EM wave passing through each meta-atom on the source encoding layer acts as a secondary source to illuminate all the meta-atoms on the first metasurface layer, and this process is repeated as the waves propagate through all metasurface layers \cite{lin2018all}. Let $\mathbf{B} = \mathbf{Z}^{L} \mathbf{W}^{L} \ldots \mathbf{Z}^{2} \mathbf{W}^{2} \mathbf{Z}^{1} \mathbf{W}^{1} \in \mathbb{C}^{N \times N}$ denote the EM response of the semantic encoding layers. Hence, the semantic encoded signal $\tilde{\mathbf{x}} \in \mathbb{C}^{N}$ can be represented as
\begin{align}
 \tilde{\mathbf{x}} = \mathbf{B}\mathbf{x} = \sqrt{p_{t}}s \mathbf{Z}^{L} \mathbf{W}^{L} \ldots \mathbf{Z}^{2} \mathbf{W}^{2} \mathbf{Z}^{1} \mathbf{W}^{1} \mathbf{Z}^{0} \mathbf{w}^{0}.
\end{align}

Note that both the source and semantic encoding involved in (3) occur naturally as the EM waves propagate through the SIM, which is in sharp contrast to the baseband semantic encoder.
\subsection{Channel Model}
The wireless channel is assumed to be block fading, denoted by $\mathbf{H} = \left[\mathbf{h}_{1}, \mathbf{h}_{2}, \ldots, \mathbf{h}_{M}\right]^T \in \mathbb{C}^{M \times N}$, where $\mathbf{h}_{m}^{T} \in \mathbb{C}^{1 \times N}$, $m \in \mathcal{M}$, denotes the channel from the output layer of SIM to the $m$-th antenna of the receiver. In this letter, we adopt the Rician fading model \cite{yao2024channel}, yielding
\begin{align}
\label{Eq4}
 \mathbf{H}=\sqrt{\frac{Kp}{1+K}}\mathbf{H}_{\mathrm{LOS}}+\sqrt{\frac{p}{1+K}}\mathbf{H}_{\mathrm{NLOS}},
\end{align}
where $p$ is the path loss, $K$ is the Rician factor, $\mathbf{H}_{\mathrm{LOS}}$ is the line-of-sight (LOS) component, characterizing the direct path between SIM and receiver, which is defined by ${[\mathbf{H}_{\mathrm{LOS}}]}_{m,n} = e^{-j\frac{2\pi}{\lambda}d_{m,n}}$, where $d_{m,n}$ denotes the distance from the $n$-th meta-atom on the output layer of the SIM to the $m$-th antenna at the receiving array. $\mathbf{H}_{\mathrm{NLOS}}$ is the non-LOS (NLOS) component, following the circularly symmetric complex Gaussian distribution with a zero mean and an identity covariance matrix. In this paper, we focus on the optimization of SIM to enable semantic encoding by assuming that the wireless channel coefficients have been estimated using advanced channel estimation approaches \cite{yao2024channel}.

As a result, the received signal can be modeled as
\begin{align}
 \label{Eq5}
 \mathbf{y} = \mathbf{H}\tilde{\mathbf{x}} + \mathbf{v},
 \end{align}
where $\mathbf{y} = \left[ y_{1}, y_{2}, \ldots, y_{M} \right] \in \mathbb{C}^{M}$ represents the signal received at the receiving array, $\mathbf{v} \in \mathbb{C}^{M}$ is the additive white Gaussian noise (AWGN), satisfying $\textbf{v}\sim\mathcal{CN}\left(\mathbf{0}, \sigma^{2}_{v}\mathbf{I_{M}}\right)$, where $\mathbf{I_{M}} \in \mathbb{C}^{M \times M}$ denotes the identity matrix and $\sigma^{2}_{v}$ represents the average noise power.

\subsection{Receiver}
As shown in Fig. 1, $M$ antennas are arranged at the receiver in a uniform linear array with spacing $d_{r}$. Each antenna corresponds to an image class. In this letter, we aim to train the SIM-aided EMNN to steer the wireless signals $\tilde{\mathbf{x}}$ towards the corresponding antenna. As a result, the image can be recognized by comparing the magnitude of the signal at each antenna. This process can be formulated as
\begin{align}
 \hat{m} = \arg \max_{m \in \mathcal{M}} { \left\{|y_{1}|^{2}, |y_{2}|^{2}, \ldots, |y_{M}|^{2} \right\}},
\end{align}
where $\hat{m}$ is the index of the antenna with maximum received energy, also indicating the corresponding class of the image.

\section{SIM-aided EMNN for Semantic Encoding}
In this section, we first introduce the loss function used for evaluating the performance of the EMNN. Then, the detailed training process of the SIM is explained.
\subsection{Loss Function}
The cross-entropy (CE) is used as the loss function to measure the difference between the power pattern across the receiving array and the expected probability distribution of recognizing the input image. Specifically, we use the softmax function to normalize the power at the receiving array, i.e., $\tilde{\mathbf{y}} = \text{softmax}(\mathbf{|y|}^{2}) = \left[ \tilde{y}_{1}, \tilde{y}_{2}, \ldots, \tilde{y}_{M} \right] \in \mathbb{R}^{M}$, where $\tilde{y}_{m} = \frac{e^{|{y_m}|^{2}}}{\sum_{i=1}^{M} e^{|{y_i}|^{2}}}$, $m \in \mathcal{M}$. Moreover, let $\mathbf{q} = \left[q_{1}, q_{2}, \ldots, q_{M} \right] \in \mathbb{R}^{M}$ denote the probability distribution, which is defined by
\begin{align}
 q_{m}=\begin{cases}1, \quad \text{$m$ is the class of the source image},\\0, \quad \text{otherwise}.\end{cases}
\end{align}

As a result, the CE is defined as
\begin{align}
\label{Eq8}
 \mathcal{L}_{\mathrm{CE}}(\mathbf{a}^{l},\bm{\phi}^{l})=-\sum_{m = 1}^{M}q_{m}\log(\tilde{y}_{m}).
\end{align}
Note that a smaller value of CE indicates that the energy distribution at the receiving array is closer to the expected probability distribution.

\subsection{Training of EMNN}
In this subsection, we aim to train the transmission coefficients, including the amplitude vectors $\{\mathbf{a}^{1}, \mathbf{a}^{2}, \ldots, \mathbf{a}^{L}\}$ and phase vectors $\{\bm{\phi}^{1}, \bm{\phi}^{2}, \ldots, \bm{\phi}^{L}\}$, of the SIM for minimizing the total loss function of all image samples in the dataset\footnote{For discrete phase shifts, one can optimize the SIM-aided EMNN using an alternating optimization approach \cite{hassan2024efficient}.}. The training process mainly includes six steps:
\begin{enumerate}
 \item Initialize the transmission coefficients of the SIM, where $a^{l}_{n}$ and $\phi^{l}_{n}$, $n \in \mathcal{N}$, $l\in\mathcal{L}/\{0\}$ obey the uniform distribution of $ \mathcal{U}\left[0, 1\right )$ and $ \mathcal{U}\left[ 0, 2\pi \right )$ respectively.
 \item Extract $B$ samples from the image dataset to create a mini-batch and preprocess each image to match the size of the source encoding layer.
 \item For each image in the mini-batch, generate a set of received signal samples according to (5).
 \item Calculate the loss function $\mathcal{L}_{\mathrm{CE}}$ according to \eqref{Eq8} for each image in the mini-batch and add them up to obtain the total loss function over the mini-batch, denoted by $\mathcal{L}_{\text{total}}$.
 \item Use the Adam optimization algorithm \cite{Kingma2014AdamAM} to iteratively update the transmission coefficients of the SIM. It is worth mentioning that both parameters $\mathbf{a}^{l}$ and $\bm{\phi}^{l}$ are updated simultaneously.
 \item At each iteration, project the SIM transmission coefficients to the feasible set by applying the MinMaxScaler normalization method \cite{SINGH2020105524Investigating}.
\end{enumerate}

After applying the mini-batch gradient descent method in several iterations, the loss function value converges. The detailed steps for training the SIM-aided EMNN to enable image recognition tasks in the wave domain are given in Algorithm~1, where $\nabla$ in step 7 represents the gradient operator. Additionally, $\alpha > 0$ in step $7$ is the learning rate, while $\mathbf{m}_{t}^{\{\mathbf{a}_{t}^{l}, \bm{\phi}_{t}^{l}\}}$ and $\mathbf{v}_{t}^{\{\mathbf{a}_{t}^{l}, \bm{\phi}_{t}^{l}\}}$ are respectively the first-order and second-order moment vectors. Additionally, $\beta_1$ and $\beta_2$ represent the decay coefficients, while $\epsilon$ is a small positive constant used for enhancing numerical stability. Appropriately adjusting these hyperparameters can strike tradeoffs between convergence speed and image recognition accuracy.

\newcommand{\INDSTATE}[1][1]{\STATE\hspace{#1\algorithmicindent}}
\begin{algorithm}[!t]
 \caption{Training of the SIM-based EMNN for enabling image recognition task-oriented SemComs.}
 \label{alg:AOA}
 \renewcommand{\algorithmicrequire}{\textbf{Input:}}
 \renewcommand{\algorithmicensure}{\textbf{Output:}}
 \begin{algorithmic}[1]
 \REQUIRE $B$ image samples, channel matrix $\mathbf{H}$, EM propagation coefficient matrices $\{\mathbf{w}^{0}, \mathbf{W}^{1}, \mathbf{W}^{2}, \ldots, \mathbf{W}^{L}\}$.
 \STATE Initialize the SIM transmission coefficients $\mathbf{a}_{t}^{{l}}$ and $\bm{\phi}_{t}^{{l}}$, and the iteration counter $t = 0$.
 \STATE Preprocess all input images from the mini-batch to match the size of the source encoding layer.
 \STATE \textbf{Repeat}
 \INDSTATE[1] Calculate $\mathbf{y}$ according to \eqref{Eq5} for each image.
 \INDSTATE[1] Calculate the loss function $\mathcal{L}_{\mathrm{CE}}\left(\mathbf{a}_{t}^{l},\bm{\phi}_{t}^{l}\right)$ according to \phantom{00} (8) for each image as well as the total loss $\mathcal{L}_{\text{total}}$ over \phantom{00} the mini-batch.
 \INDSTATE[1] Update $t \leftarrow t+1$.
 \INDSTATE[1] Update SIM transmission coefficients simultaneously \hspace*{1.1em} by leveraging the Adam optimizer:
\begin{scriptsize}\begin{align}
 \mathbf{g}_{t}^{\{\mathbf{a}_{t}^{l},\bm{\phi}_{t}^{l}\}} &= \nabla_{\{{\mathbf{a}_{t-1}^{{l}},\bm{\phi}_{t-1}^{{l}}}\}}\mathcal{L}_{\text{total}}(\mathbf{a}_{t-1}^{{l}},\bm{\phi}_{t-1}^{{l}}),\notag\\
 \mathbf{m}_{t}^{\{\mathbf{a}_{t}^{l},\bm{\phi}_{t}^{l}\}}&=\beta_{1} \mathbf{m}_{t-1}^{\{\mathbf{a}_{t-1}^{l},\bm{\phi}_{t-1}^{l}\}}+(1-\beta_{1}) \mathbf{g}_{t}^{\{\mathbf{a}_{t}^{l},\bm{\phi}_{t}^{l}\}},\notag\\
 \mathbf{v}_{t}^{\{\mathbf{a}_{t}^{l},\bm{\phi}_{t}^{l}\}}&=\beta_{2}\mathbf{v}_{t-1}^{\{\mathbf{a}_{t-1}^{l},\bm{\phi}_{t-1}^{l}\}}+(1-\beta_{2}) \left(\mathbf{g}_{t}^{\{\mathbf{a}_{t}^{l},\bm{\phi}_{t}^{l}\}} \circ \mathbf{g}_{t}^{\{\mathbf{a}_{t}^{l},\bm{\phi}_{t}^{l}\}}\right), \notag\\
 \hat{\mathbf{m}}_{t}^{\{\mathbf{a}_{t}^{l},\bm{\phi}_{t}^{l}\}}& =\frac{\mathbf{m}_{t}^{\{\mathbf{a}_{t}^{l},\bm{\phi}_{t}^{l}\}}}{1-\beta_{1}^{t}},\mathbf{\hat{v}}_{t}^{\{\mathbf{a}_{t}^{l},\bm{\phi}_{t}^{l}\}} =\frac{\mathbf{v}_{t}^{\{\mathbf{a}_{t}^{l},\bm{\phi}_{t}^{l}\}}}{1-\beta_{2}^{t}},\notag\\
 \{\mathbf{a}_{t}^{l},\bm{\phi}_{t}^{l}\} &=\{\mathbf{a}_{t-1}^{l},\bm{\phi}_{t-1}^{l}\}-\alpha\frac{\hat{\mathbf{m}}_{t}^{\{{\mathbf{a}_{t}^{l}}, \bm{\phi}_{t}^{l}\}}}{\sqrt{\hat{\mathbf{v}}_{t}^{\{{\mathbf{a}_{t}^{l}}, \bm{\phi}_{t}^{l}\}}+\epsilon}}.\notag
 \end{align}\end{scriptsize}
 \INDSTATE[1] Project $\mathbf{a}_{t}^{{l}}$ and $\bm{\phi}_{t}^{{l}}$ to the feasible set by applying the \phantom{00} MinMaxScaler method.
 \STATE \textbf{Until} The total loss $\mathcal{L}_{\text{total}}$ achieves convergence or the maximum number of iterations is achieved.
 \ENSURE $\{\mathbf{a}^{1}, \mathbf{a}^{2}, \ldots, \mathbf{a}^{L}\}$, and $\{\bm{\phi}^{1}, \bm{\phi}^{2}, \ldots, \bm{\phi}^{L}\}$. 
 \end{algorithmic}
\end{algorithm}
\section{Simulation Results}
\subsection{Simulation Settings}
In this section, we use the MNIST dataset to evaluate the performance of the proposed SIM-based EMNN for implementing image recognition tasks in the wave domain. It consists of approximately 70,000 images of handwritten digits from 0 to 9 with $28 \times 28$ grayscale pixels. The cardinality ratio of the training set to the testing set is 6:1, and 20\% of the training set is allocated for validation purposes. Note that the noise is added to the received signal during the training process to improve the system's robustness. Moreover, the SemCom system operates at $28$ GHz, corresponding to a wavelength of $\lambda = 10.7$ mm. The thickness of the SIM is set to $D = 10\lambda$, while the spacing between adjacent meta-atoms is a wavelength, yielding $S_{a} = \lambda^{2}$. Additionally, the horizontal distance from the transmitting antenna to the input layer of the SIM is set to $d_{t,s} = d_{s}$. The distance between adjacent receiving antennas is set to $d_r = \lambda/2$. For the wireless link, the Rician factor is set to $K = 3$ dB, and the path loss from the SIM to the receiver is modeled as $p = C_0d_{s,r}^{-\gamma}$, where $C_{0} = -35$ dB denotes the path loss at a reference distance of 1 m. $d_{s,r}$ denotes the distance between the SIM and the receiving array, and $\gamma =2.8$ is the path loss exponent in our simulations \cite{an2021low}. Additionally, the transmit power and the noise power are set to $p_{t} = 40$ dBm and $\sigma^{2}_{v} = -104$ dBm, respectively.

In terms of the training process of SIM-based EMNN, we consider 100 epochs, and the mini-batch size $B$ is set to $64$. Moreover, the learning rate is set to $\alpha = 0.001$, which will be gradually decreased by multiplying $\alpha$ with a decay factor $\iota_{f} = 0.8$, if the loss function value does not vary in five consecutive epochs. The decay coefficients of the first-order and the second-order moment estimates are $\beta_{1} = 0.9$ and $\beta_{2} = 0.999$, respectively. The first-order moment vector $\mathbf{m}_{t}$ and the second-order moment vector $\mathbf{v}_{t}$ are initialized to $\mathbf{0}$.

\begin{figure}[!t] 
\centering
\subfloat[\scriptsize Recognition accuracy under different numbers of metasurface layers.] 
{\label{Figure21}\includegraphics[width=6.5cm]{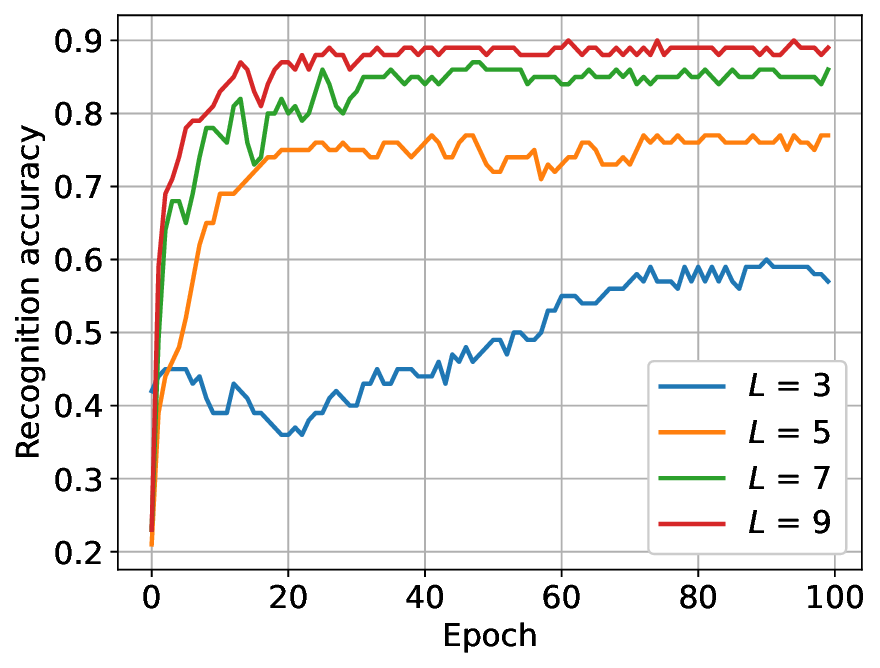}}\
\subfloat[\scriptsize Recognition accuracy under different numbers of meta-atoms on each layer.]
{\label{Figure22}\includegraphics[width=6.5cm]{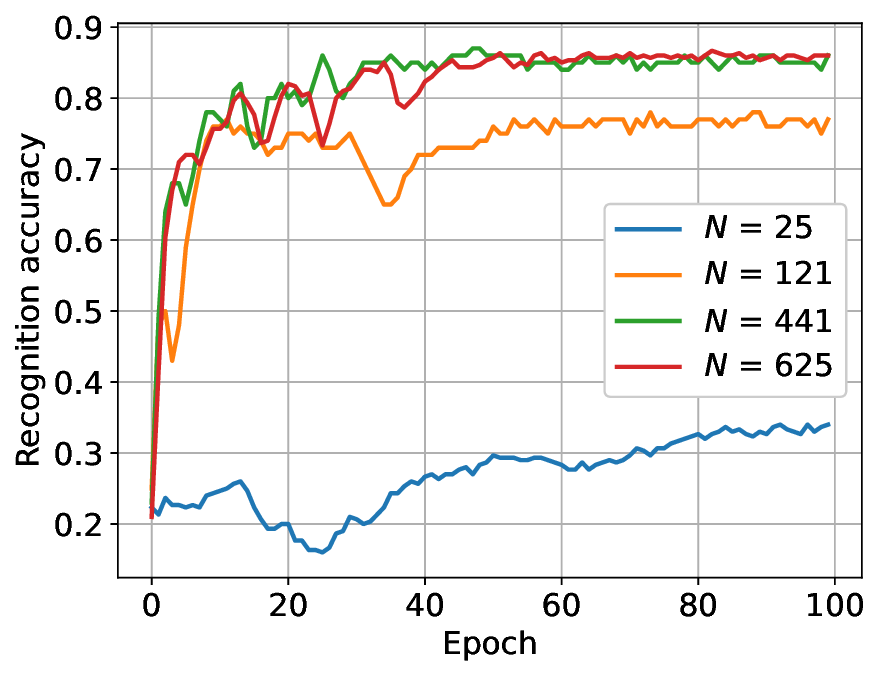}}
\\ 
\subfloat[\scriptsize Recognition accuracy under different propagation distances.] 
{\label{Figure23}\includegraphics[width=6.5cm]{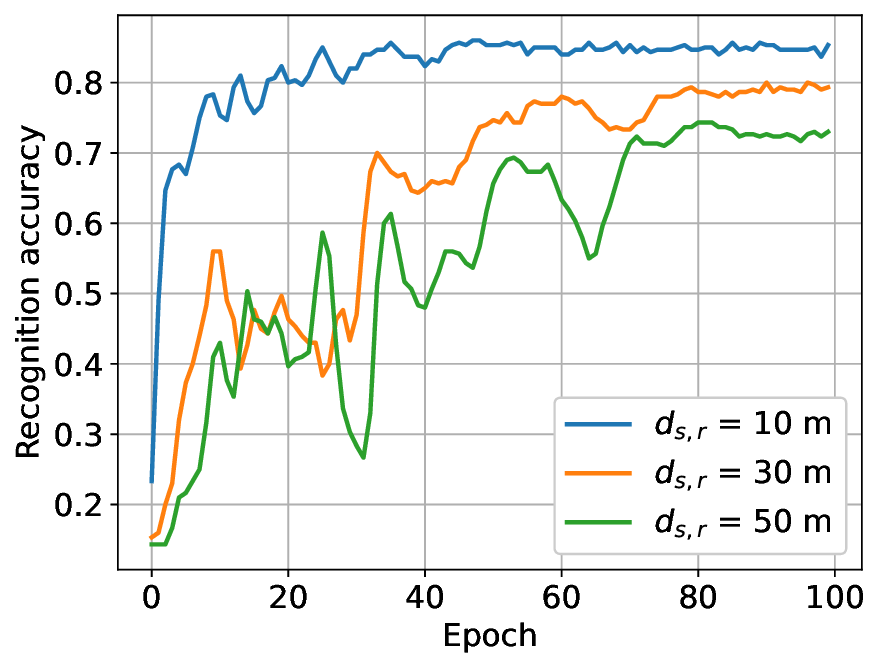}}\
\subfloat[\scriptsize Recognition accuracy under different numbers of image classes.]
{\label{Figure24}\includegraphics[width=6.5cm]{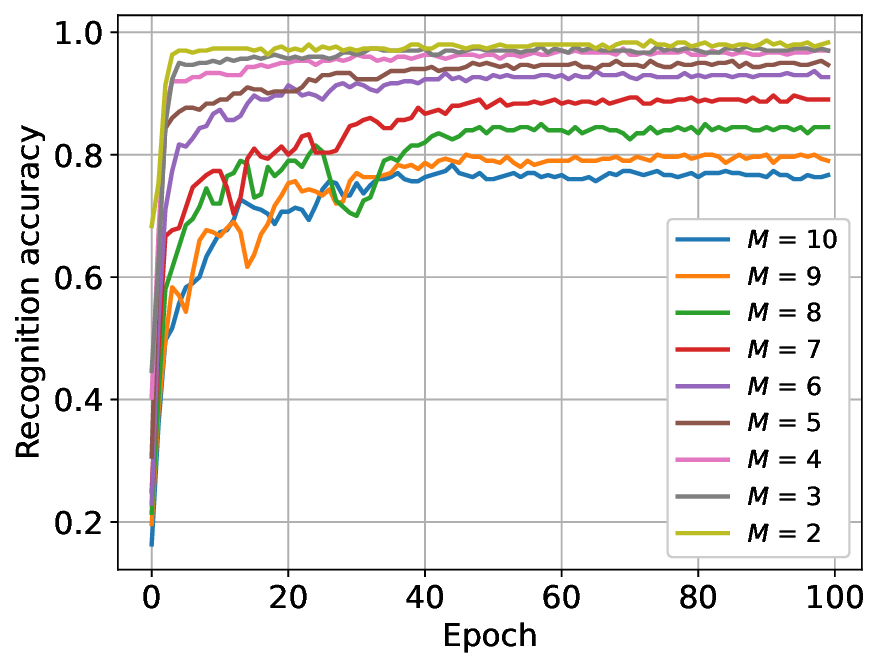}} 
\caption{Recognition accuracy using the SIM-based EMNN.}
\label{fig_2} 
\end{figure}
\subsection{Performance Evaluation}
In Fig. 2(a), we evaluate the effect of the number $L$ of layers on the recognition accuracy, where we consider $N = 441$, $d_{s,r} = 10$ m. For each input image, we retain only the $21 \times 21$ pixels in the center. For simplicity, $M = 8$ antennas are deployed at the receiver for recognizing digits $0 \sim 7$. As observed in Fig. 2(a), the recognition accuracy using the SIM-based EMNN increases with the number of metasurface layers, thanks to the enhanced inference capability of the multi-layer architecture for achieving more accurate beam steering. 

Fig. 2(b) shows the recognition accuracy under different numbers $N$ of meta-atoms on each layer, where we set $L = 7$ and maintain the other parameters the same as Fig. 2(a). For each input image, we preprocess it to match the size of the source encoding layer by using the bilinear interpolation method \cite{jahne2005digital}. Note that the recognition accuracy improves as $N$ increases, since a larger number of learnable neurons (meta-atoms) would enhance the representational capability of the SIM. Additionally, increasing the number of meta-atoms also increases the aperture of the output metasurface layer, thus having a higher Rayleigh resolution to distinguish the signals at different receiving antennas. Specifically, the seven-layer SIM with $N = 25$ meta-atoms on each layer achieves a low recognition accuracy of only $32\%$ on the testing set. As $N$ increases to $121$ and $441$, the recognition accuracy on the testing set becomes $77\%$ and $84\%$, respectively. 

According to the simulation results in Fig. 2(a)-(b), increasing the value of $L$ and $N$ enhances recognition accuracy. However, the improvement of recognition accuracy is relatively small when $L$ exceeds $7$, as shown in Fig. 2(a). Similarly, Fig. 2(b) demonstrates the performance plateaus when $N$ exceeds 441. These diminishing returns suggest choosing proper SIM parameters for achieving the desired performance with minimum hardware cost.

Fig. 2(c) illustrates the recognition accuracy versus the training epoch as the distance $d_{s,r}$ between the SIM and receiver increases, with $N = 441$ and other parameters unchanged. It is observed from Fig. 2(c) that a shorter propagation distance would improve the recognition accuracy. This is due to a smaller value of $d_{s,r}$ would cause stronger channel gain compared to that with a longer propagation distance. Hence, the noise effects would become increasingly negligible. Secondly, as the receiving array moves closer to the transmitter, the channels from the SIM to different antennas would become more distinguishable. As a result, decreasing the propagation distance from $d_{s,r} = 50$ m to $d_{s,r} = 10$ m enhances the recognition accuracy from $72\%$ to $84\%$.

Fig. 2(d) illustrates the recognition accuracy under different numbers of image classes where we set $d_{s,r} = 10$ m. Specifically, we examine nine scenarios by gradually increasing both the number of image classes as well as the number of receiving antennas. Note that recognizing more handwritten digits requires the SIM to accurately focus multiple beams on an increased number of receiving antennas. As observed from Fig. 2(d), the recognition accuracy becomes worse as the number of image classes increases. Additionally, a faster convergence trend is observed as $M$ decreases.

\begin{figure}[!t]
\centering
\includegraphics[width=0.4\textwidth]{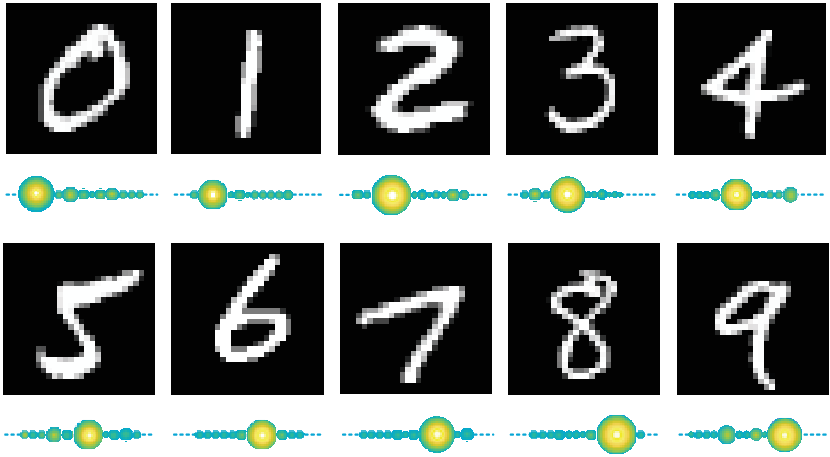}
\caption{Input image and the corresponding energy distribution at the receiving array.}
\label{fig_3}
\end{figure} 

\begin{figure}[!t]
\centering
\includegraphics[width=0.4\textwidth]{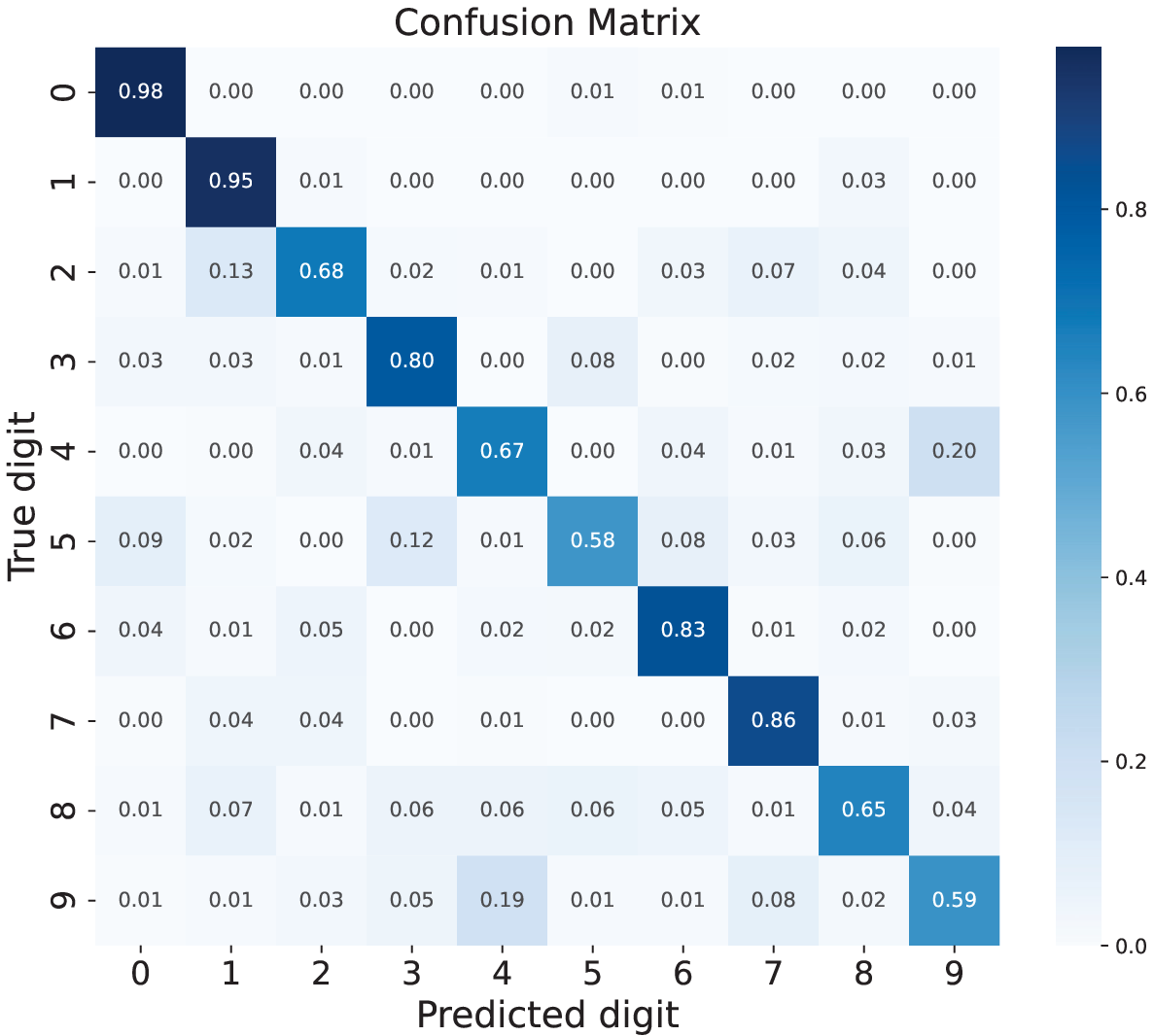}
\caption{The confusion matrix for recognizing ten digits.}
\label{fig_4}
\end{figure} 

Finally, to better illustrate the effectiveness of the proposed SIM-based EMNN, Fig. 3 depicts the input image and the corresponding energy distribution at the receiving array, while the confusion matrix of utilizing the SIM-based EMNN to recognize ten handwritten digits is shown in Fig. 4. It is demonstrated that the EMNN is capable of focusing the information-bearing signals passing through the SIM onto the corresponding antenna.
\section{Conclusion}
In this letter, we investigated a SIM-aided image recognition task-oriented SemCom. We trained a SIM in front of the transmitting antenna to steer the signals passing from the image source encoding layer towards the corresponding antenna. As a result, the receiver identified the image class by probing the magnitude of the received signal. Extensive simulation results verify the effectiveness of the SIM-based EMNN in achieving over 90\% recognition accuracy for image recognition task-oriented SemComs. Combined with traditional modulated signals, SIM is expected to further achieve multimodal semantic information transmission.

\bibliographystyle{IEEEtran}
\bibliography{IEEEabrv,Ref}
\end{document}